# Register Spilling for Specific Application Domains in ASIPs

M.G.G.C.R. Salgado
Department of Statistics & Computer Science
University of Peradeniya
Peradeniya, Sri Lanka
chaluka.salgado@gmail.com

R.G. Ragel
Department of Computer Engineering
University of Peradeniya
Peradeniya, Sri Lanka
roshanr@pdn.ac.lk

*Abstract* — An Application Specific Instruction set Processor (ASIP) is an important component in designing embedded systems. One of the problems in designing an instruction set for such processors is determining the number of registers is needed in the processor that will optimize the computational time and the cost. The performance of a processor may fall short due to register spilling, which is caused by the lack of available registers in a processor. In the design perspective, it will result in processors with great performance and low power consumption if we can avoid register spilling by deciding a value for the number of registers needed in an ASIP. However, as of now, it has not clearly been recognized how the number of registers changes with different application domains. In this paper, we evaluated whether different application domains have any significant effect on register spilling and therefore the performance of a processor so that we could use different number of registers when building ASIPs for different application domains rather than using a constant set of registers. Such utilization of registers will result in processors with high performance, low cost and low power consumption.

## I. Introduction

An Application Specific Instruction-set Processor (ASIP) [8] is a programmable processor, which is designed to perform a specific task more efficiently compared to a general purpose processor. This extra efficiency is not only associated with a higher performance, but also other factors such as low manufacturing cost and low power consumption. These processors are being more popularly used in embedded devices to satisfy performance requirements of today's world. ASIPs allow designers to customize both the Instruction Set Architecture (ISA) and the underlying microarchitecture for a specific application domain. An ASIP instruction set is usually different than a generic instruction set. It does not have to be composed of a mnemonic and register/memory operands. In a typical RISC processor, instructions trigger functional units along with general purpose register addresses. However, ASIPs can benefit also from configuration registers or utilizes specially designed data flow mechanisms that are hardwired in the system. Both high flexibility, application-oriented high performance and energy efficiency can be achieved in ASIPs.

Registers are a very small amount of very fast memory that is built into the CPU (Central Processing Unit) in order to speed up its operations by providing quick access to commonly used values. Memory refers to semiconductor device whose content can be accessed (read and written to) at high speeds but which are held there only temporarily (while in use or only as long as the power supply remains on). Most memory consists of main memory which is comprised of RAM (Random Access Memory) chips that are connected to the CPU by a bus.

When a program is executed, the CPU needs to store its live variables in its registers. However, each CPU has a limited number of registers, which are used by every function it executes. It is possible that some functions can share the same registers. Therefore, those registers may be overwritten so that some programs may lose their data. We can keep important registers from being overwritten by another function by saving them to the main memory before the function executes and then restore them back when we need them and such process is called *register spilling* [6]. Register spilling happens when the number of live variables during the execution of a program is greater than the number of available registers. Due to the unavailability of registers, their values are saved into a specific location in the main memory called the stack, where the values can be kept safe until they are needed later by the program.

RISC processors typically need more memory than CISC do to store the same program. Except in the most speed-critical of embedded devices, the cost of memory is much more critical than the execution speed of the processor. To reduce memory requirements and, thereby, cost, Advanced RISC Machines (ARM) created the THUMB instruction set as an option for their RISC processor cores. The most well-known chip that includes the THUMB instruction set is the ARM7TDMI [5]. The THUMB instruction set consists of 16-bit instructions that act as a compact shorthand for a subset of the 32-bit instructions of the standard ARM. Every THUMB instruction could instead be executed via the equivalent 32-bit ARM instruction. However, not all ARM instructions are available in the THUMB subset. In THUMB state registers from r8 to r12 are called *special registers*, which are not used to execute THUMB instructions. All the other registers (r0-r7 and r13-r15) of ARM7TDMI are used in THUMB state and are called regular registers. In this study we have performed our experiments on this ARM7TDMI processor in THUMB state.

In this paper, we have performed a study using the ARM THUMB processor and a set of domain specific applications to address a few problems such as whether a specific set of application domains has any effect on register spilling, whether it is possible to decide a value for the number of additional registers required to stop register spilling and also whether that value varies through application domains.

The remainder of the paper is organized as follows. Section 2 gives a brief description of the literature relevant to the problem. Section 3 describes the proposed spills and additional registers count approach and its implementation. Section 4 presents experimental results, followed by conclusion and future work by Section 5.

## II. LITERATURE REVIEW

Fraser and Hanson [1] simplified spill manager to spill the registers whose next use is the most distant. In their paper they describe the management of register spills in a retarget-able C compiler. The trade-offs have been arranged so that the common case (no spills) generates respectable code quickly and the uncommon case (spills) is less efficient but as simple as possible. Still it causes more time in execution because the code consist spills itself though the compiler spills the register with less demand. In 1982 G. J. Chaitin discovered [2] how to extend the graph coloring approach so that it naturally solves the spilling problem. Previously the compiler produced the spill code so the quality of the spill code was much low and amount of time taken is considerably high. In their study, spill decisions are then made on the basis of the register conflict graph and cost estimates of the value of keeping the result of a computation in a register rather than in the storage. According to their experiments now register allocation technique takes the better advantage of speed potential of using registers to store than previous approaches. Andrew W. Appeal and Lal George [3] suggested a register allocation algorithm for machines with few registers. According to their study register allocation by graph coloring has been successful in machines with 30 or more registers. But it became worst when it is used in machines with few registers because of higher register spilling.

A hardware/software cooperative approach and a linear scan register allocation algorithm to utilize the existing custom registers in ASIPs for eliminating register spills was proposed by Lin and Fei [4]. The data traffic between the processor and memory can be reduced through efficient on-chip communications between the base processor core and custom hardware extensions. Their experimental results demonstrate that a promising performance gain can be achieved, which is orthogonal to improvements by any other technique in ASIP design. They have proposed a novel approach to turn the custom registers to a register file extension, so that the base instructions can be used to store the data in custom registers instead of going to memory system, possibly reducing the memory traffic and hence execution time and energy consumption. According to their conclusion the limited on-chip data storage resources in ASIPs have become a major performance bottleneck. The large traffic between processor registers and the main memory results in degrading performance and increasing energy consumption.

In this paper, we prove that a number of additional registers can be used to avoid register spilling and the number of additional registers needed varies with the application domain. Additional registers are the ones which we can add to the original set of registers of the processor to increase the number of registers available. In our study, first we looked over the spilling values to check whether it has any effect on the application domain and also whether it is possible to stop register spilling using a set of additional registers. Finally we tried to determine a specific value for number of additional registers needed for each application domain.

## III. SOFTWARE IMPLEMENTATION

### A. Spill Count Approach

In our approach to count spills, we considered both number of spills and the overhead due to such. We have performed all our calculations using the static assembly code of each application we used. Therefore those values will be the representative spills as opposed to the exact impact that should be calculated using dynamic instructions of the programs.

Fig.1 shows the spills we have considered for our experiment. As it can be seen, initially (I1), some useful register values are pushed onto the stack for the further usage. This is common to each and every function in a program. In both I2 and I3, we can see that two special register values are moved into regular registers and then in the I4, values of those regular registers are pushed onto the stack. We have considered these movements as spill overheads as it is performed to have free spaces for spills that are about to happen. Therefore now compiler can use those two special registers to store variables. I5 and I6 show spills where the values of regular registers are moved into special registers.

Meanwhile the compiler uses those two regular registers to store other two live variables, and whenever the compiler needs the values in special registers, it will move them to regular

```
I1:     push    {r4,r5,r6,r7,lr}
I2:     mov     r1,r9
I3:     mov     r3,sl
I4:     push    {r1,r3}
        ....
        ....
        ....
I5:     mov     r9,r2
I6:     mov     sl,r0
        ....
        ....
        ....
I7:     mov     r3,sl
I8:     mov     r1,r9
        ....
        ....
I9:     pop     {r2,r3}
I10:    mov     r9,r2
I11:    mov     sl,r3
I12:    pop     {r4,r5,r6,r7,pc}
```

Fig. 1. Spill Count Instructions Example

registers and use them for calculations as the I7, I8 show. Then the values which have been pushed onto the stack will be popped back to the regular registers and then again move to special registers just like I10 and I11 show. Those steps have also been considered as spill overheads. So the summation of both spills and spill overheads has been considered as the total spill value of each program.

This experiment has been conducted considering few programs in different application domains from an embedded system benchmark suite [7]. Each program has its own uniqueness in execution. Since an application is a collection of different functions, the summation of spill values for each function was considered in determining the final spilling value of each application. Our primary task was to check whether the total spills for each application varies with its application domain. Meanwhile, we tried to find a value for additional registers set which will avoid spilling in those applications.

*B. Additional Register Count Approach*

To avoid the register spilling the processor should have enough registers to store live variables. As we discussed earlier, spilling happens when a processor runs out of free registers to store program variables. So we can decide how many registers it needs by considering available live variables at each execution of the functions of the program. To count the additional registers, we used an algorithm which uses the similar concept we used to calculate spilling values. As the Fig. 2 shows there we assumed that when a regular register value is moved into a special register (a spill), means that the compiler doesn't have enough registers for live variables. So that we considered a value which increases when a register spill happens and decreases with a reload as the Fig. 2 shows. At the end of the function we could have a value which is the number of additional registers we need to avoid spilling for that function. Likewise we can have a set of values for additional registers for each function. Then we can decide a value for the program by taking the maximum value from the set of values. Spill overheads were ignored and only the spills have been considered to count the additional registers set.

To overcome the problem that some registers are reloaded multiple times, a list was used to stores the target special register of the spilled register (i.e. for mov sl, r0 then sl is added to the list). And when the restore happens, it removes the register involved with that reload from the register list (i.e. for mov r1, sl then sl is removed from the list). When a reload occurs, the program check whether the special register involved with that instruction is in the register list and then decides whether to decrease the value of the number of additional registers to avoid unnecessary decrements.

*C. Implementation of the above Approaches*

We proposed spills and additional register count algorithm aligned with the above two approaches. The pseudo code of the algorithm is given in Algorithm 1. First we read instructions one at a time and checked the type of instruction, whether a push, a pop or a mov. When a push or a pop instruction occurs, we increase the overhead count by the number of registers involved in that instruction. Since the first push and the last pop instructions of each function of the assembly code are not involved with register spilling, we should not increase the overhead count for those instructions. To prevent such unnecessary increments, we maintained a value called ***check***. And the overhead count is increased only if the ***check*** value is greater than zero.

When it is a mov instruction, first, we check whether at least one of the two registers involved with that instruction is a special register. When one of those registers is a special register, then that instruction could be either a spill or a spill overhead. Therefore, with the help of the ***check*** variable, we can identify whether it is a spill overhead (when ***check*** value is equal to 1) or a spill (otherwise).

When the first register is a special register, if it is a spill overhead, then we increase the overhead count and decrease the additional registers count. Otherwise, it is a spill and therefore we increase both spill count and additional registers count. Meanwhile, we add that register to an array called ***reg_in_use***, which stores the set of registers that holds spilled register values. When the second register is a special register, if it is a spill overhead, then we increase both additional registers count and the overhead count values. Otherwise, we increase the spill count and check whether that register already exists in the ***reg_in_use*** array and if it does we remove the register from the array and decrease the additional register count.

```
I1:    push   {r4,r5,r6,r7,lr}
I2:    mov    r1,r9
I3:    mov    r3,sl
I4:    push   {r1,r3}
       ....
       ....
I5:    mov    r9,r2            <- RegList: r9       RegCount=1
I6:    mov    sl,r0            <- RegList: r9,sl    RegCount=2
       ....
       ....
I7:    mov    r3,sl            <- RegList: r9       RegCount=1
       ....
       ....
I8:    mov    r1,sl            <- RegList: r9       RegCount=1
       ....
       ....
I9:    pop    {r2,r3}
I10:   mov    r9,r2
I11:   mov    sl,r3
I12:   pop    {r4,r5,r6,r7,pc}
```

Fig. 2. Spill Count Instructions Example

**Algorithm 1: Spills and Additional Register Count**

**INPUT:**
{ check } is the variable used to overcome unnecessary push/pop instructions
{ spill_count } is the total number of spills
{ overhead_count } is the total number of spill overheads
{ add_reg_count } is the total number of additional registers we need to stop spilling
{ reg_in_use } is an array that stores special registers which hold spilled register values
{ $R_1$ } is the first register of the mov instruction
{ $R_2$ } is the second register of the mov instruction
/* eg. mov $R_1,R_2$ */

```
BEGIN

WHILE no instruction to read DO

Read assembly code file instruction by instruction.

IF a push instruction THEN
    check++
    IF check>1 THEN
        Add no of registers involved to overhead_count
    END IF

ELSE IF a pop instruction THEN
    IF check>1 THEN
        Add no of registers involved to overhead_count
    END IF
    check --

ELSE IF a mov instruction THEN

    IF R1 OR R2 are special registers THEN
        IF R1 is a special rgister THEN
            IF check==1 THEN
                overhead_count++
                add_reg_count--
            ELSE
                add_reg_count++
                spill_count++
                Add R1 register to reg_in_use
            END IF

        ELSE IF R2 is a special rgister THEN
            IF check==1 THEN
                add_reg_count++
                overhead_count++
            ELSE
                spill_count++
                IF R2 in reg_in_use THEN
                  Remove that register from the list
                    add_reg_count--
                END IF
            END IF
        END IF
    END IF

END IF

END WHILE
```

According to Algorithm 1, we implemented an application using Java which takes an assembly file as the input and outputs the number of spills, overheads and total instructions and also the number of additional registers for each function as well as for the entire program, we used it to obtain our experimental results.

## IV. EXPERIMENTAL RESULTS

Here we have used benchmarks from Mibench [7] which is composed of freely available source code. The selected, benchmarks represent a variety of widely used embedded applications in different fields, such as network, security consumer, office and automotive. In this study we have considered those fields as application domains and from each domain we have taken at least one application.

Spilling Rate = (Spills+Overheads) / Total Instructions   (1)

First, those applications were compiled and converted into assembly code and then the number of spills, the spill overheads and the number of additional registers were counted for each application using Algorithm 1.

After analyzing the recorded, data we calculated the spilling rate for each application using (1) where summation of total spills and overheads for each application is divided by the total number of instructions of that application. Then we took the average values for each domain and tabulated the Table I. As Fig. 3 shows, then we drew a bar chart using those averages of spilling rates. There you can see a reference line which was obtained by taking the average of spilling rates of all the application domains which was approximately 0.1195.

There were three application domains which were higher than the average value. They were Automotive, Network and Office. That means they have more spilling rates than the other applications due to a large number of live variables they are dealing with at their run time.

So it is obvious that they have to work with large sets of live variables which results in a higher number of spills. Even though consumers and security domains are also working with large sets of data, they are only dealing with a low number of live variables so the spilling values are less than the other domains. This clearly illustrate that there is no relation between an application's variables and spills, but the number of spills always depends on live variables of that application.

TABLE I.    AVERAGES OF SPILLING RATES

| Application Domain | Application Name | Spilling Rate | Average |
|---|---|---|---|
| automotive | bacismath | 0.111 | 0.140 |
|  | bitcnts | 0.168 |  |
|  | qsort | 0.205 |  |
|  | susan | 0.078 |  |
| network | dijkstra | 0.134 | 0.165 |
|  | patricia | 0.195 |  |
| consumer | prg2lout | 0.095 | 0.083 |
|  | lame | 0.105 |  |
|  | tiffdither | 0.070 |  |
|  | tiffmedian | 0.086 |  |
|  | tiff2bw | 0.075 |  |
|  | tiff2rgba | 0.066 |  |
| security | blowfish | 0.083 | 0.113 |
|  | pgp | 0.077 |  |
|  | rijndael | 0.128 |  |
|  | sha | 0.165 |  |
| office | stringsearch | 0.190 | 0.190 |

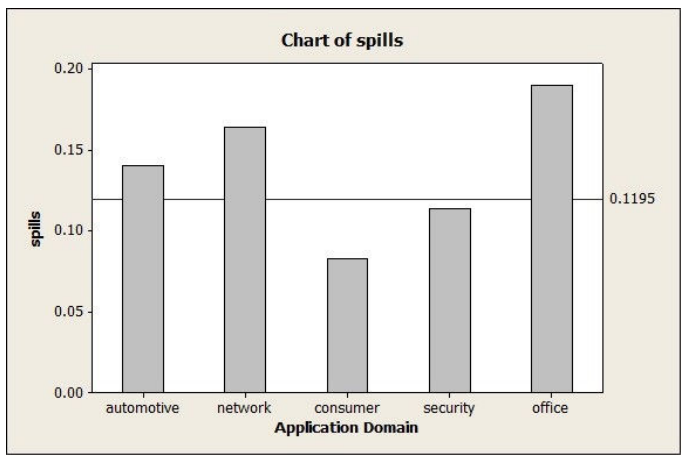

Fig. 3. Variance of the Spilling Rates

One of the important factors regarding designing an application specific processor is to decide on a number of registers. The number of additional registers needed to avoid spilling is therefore important in the designing perspective. Whether we can decide on this number based on application domains is one of our major goals. Therefore, we tried to figure it out using our data set as Table II shows. We consider the additional register counts of each application and then calculated the averages for each application domain. Then we checked whether we can decide on that value. However, there were some outliers which should be ignored to obtain a reasonable result. So we considered the data set without those outliers and did the calculations. Here we have rounded up the values to the highest nearest integer.

Table III shows average values of additional registers [omitting the outlier "40" observation] and their rounded values for each application domain. Now we can decide on a value for the number of additional registers of an application by using those rounded values according to their application domain. Now a designer can determine a specific value for the number of registers of an Application Specific Instruction set Processor using the results of our study instead of adding a constant set of registers for each ASIP.

TABLE II. AVERAGES OF ADDITIONAL REGISTERS

| Application Domain | Application Name | Additional Registers | Average |
|---|---|---|---|
| automotive | bacismath | 3 | 5.5 |
| | bitcnts | 4 | |
| | qsort | 4 | |
| | susan | 11 | |
| network | dijkstra | 4 | 5.0 |
| | patricia | 6 | |
| consumer | prg2lout | 40 | 14.3 |
| | lame | 8 | |
| | tiffdither | 6 | |
| | tiffmedian | 23 | |
| | tiff2bw | 6 | |
| | tiff2rgba | 3 | |
| security | blowfish | 7 | 9.3 |
| | pgp | 13 | |
| | rijndael | 8 | |
| | sha | 9 | |
| office | stringsearch | 3 | 3 |

TABLE III. ROUNDED VALUES OF ADDITIONAL REGISTERS

| Application Domain | Average Value | Rounded Value |
|---|---|---|
| automotive | 5.5 | 6 |
| consumer | 9.2 | 10 |
| network | 5 | 5 |
| office | 3 | 3 |
| security | 9.25 | 10 |

## V. CONCLUSIONS AND FUTURE WORK

We have formulated register spilling problem for static assembly code using ARM THUMB instruction set and obtained results for different kind of application domains. According to the results we have obtained, it is clearly convinced that the number of variables does not affect the number of register spills but the number of live variables does. And also we have concluded that the spilling value varies with the application domain and it is possible to stop register spilling for any application by using a specific set of additional registers according to their application domain.

Our main purpose was to avoid register spilling and make the program execution more efficient. For that we can propose different number of additional registers for each and every application domain rather than using a constant set for every domain. But it is hard to determine a value for each application domain. So the number of additional register values we have suggested here can be proved using customized processors, is the future work.